\documentclass[preprint,12pt]{elsarticle}

\usepackage{graphics}
\usepackage{amssymb}
\usepackage{color}

\journal{Chaos, Solitons $\&$ Fractals}

\begin{document}

\begin{frontmatter}

\title{Noise induced enhancement of\\ network reciprocity in social dilemmas}

\author{Gui-Qing Zhang$^{a}$$^{\ddagger}$, Qi-Bo Sun$^b$$^{\dagger}$, Lin Wang$^{c}$$^{\ast}$}

\address{$^a$ Department of Physics, College of Science, Nanjing Forestry
University,\\ Nanjing 210037, China\\
$^{b}$ Department of Resource and Environmental Science, School of Agriculture and Biology, Shanghai
Jiao Tong University, Shanghai 200240, China\\
$^{c}$ Adaptive Networks and Control Laboratory, Department of Electronic Engineering, Fudan
University, Shanghai 200433, China\\
{E-mail: $^{\dagger}$sunqibo1210@gmail.com; $^{\ast}$fdlwang@gmail.com}; $^{\ddagger}$nkzhanggq@163.com;}

\begin{abstract}
The network reciprocity is an important dynamic rule fostering the emergence of cooperation among selfish individuals.
This was reported firstly in the seminal work of Nowak and May, where individuals were arranged on the regular lattice
network, and played the prisoner's dilemma game (PDG). In the standard PDG, one often assumes that the players have
perfect rationality. However, in reality, we human are far from rational agents, as we often make mistakes, and behave
irrationally. Accordingly, in this work, we introduce the element of noise into the measurement of fitness, which is
determined by the parameter $\alpha$ controlling the degree of noise. The considered noise-induced mechanism remarkably
promotes the behavior of cooperation, which may be conducive to interpret the emergence of cooperation
within the population.

\end{abstract}


\end{frontmatter}

\section{Introduction}

The emergence and maintenance of cooperation among selfish individuals is a ubiquitous phenomenon extensively presenting
in various complex systems, from human or animal societies to self-replicating chemical or biological systems \cite{Axelrod84,Smith95}.
To explain this question, the evolutionary game theory creates a universal theoretical framework that has been widely
studied by diverse disciplines over the past decades \cite{Nowak06,PS10Biosystem,WZJTB,PGSFM13INTERFACE}. In particular,
the evolutionary prisoner's dilemma game (PDG) has attracted numerous attention from broad theoretical issues to specific
experimental ones, as it reflects the social conflict among the independent and selfish people in a simple but accurate
way \cite{VA01PRE,SKTHI12EPL,STWKHI12PRE,RLW07PRE,GFRTCSM12PNAS,WSP12SR,SWP12SR2}.

In a typical PDG, two involved players simultaneously make the choice: cooperate or defect. Each of them will receive a
reward $R$ if they both cooperate, and a punishment $P$ if both defect. If a player defects while the opponent chooses
to cooperate, the former one receives a temptation $T$, while the latter loser receives a sucker's payoff $S$. The ranking
of these four payoffs is $T>R>P>S$. This implies that players are more prone to defect if they both wish to maximize their
own payoff, regardless of the opponent's decision. The resulting is a social dilemma inducing the widespread defection,
which is however inconsistent with the fact that the cooperative and altruistic behavior is widely observable in reality.
To answer this puzzle, multifarious mechanisms have been proposed, such as reward and punishment \cite{re1,re2,re3,re4,re5},
voluntary participation \cite{vo1,vo2}, spatially structured population \cite{sp1,sp2,sp3,sp4,sp5,noi2,sp6,sp7,sp8,sp9,sp10,
sp11,sp12,sp13,sp14,sp15}, heterogeneity or diversity \cite{he1}, the mobility of players \cite{mo1,mo2,mo3,mo4}, age structure
\cite{ag1,ag2,ag3}, to name but a few.

The network reciprocity is a well-known dynamical rule that fosters the prevalence of cooperation \cite{Nowak06Science}.
It says that if the game players are arranged on a network, where the individuals occupy the network nodes and the links
determine who interacts with whom, the cooperators can shape compact clusters to prevent the invasion of defectors
\cite{GF07PR,spa1,spa2,spa3}. Thus the number of cooperators will preserve at a high level. Nowak and May \cite{NM92NATURE}
proposed the first model of the networking PDG, where the players were located on the square lattices. At each round of the
dynamical process, players first gathered their own payoffs via the neighboring interactions based on the regulation of PDG.
Then per player had a chance to adopt the strategy of his neighbors, if they had higher fitness. By simply
introducing the interaction structure into the consideration, the cooperative behavior can be prevalent.

%
%

In the standard PDG, it is assumed that the players have perfect rationality. However, from the perspective of behavior economics, we human
are far from rational calculators, as we often make mistakes, and behave irrationally \cite{WHEB96JEP,FB11JSE}. Therefore, at each round of
the game, players may not have the ability to obtain their payoff exactly, due to the presence of noise deriving from the fact such as the
error of observation. By concerning the diversity or discrepancy of people \cite{LCC10EPJB,WLZZZ11PLOSONE,WZC08EPL,XWSMM13PA}, the degree of
noise for different players may have some heterogeneity. With the above arguments, we extend the traditional PDG by introducing the element
of noise into the definition of fitness of individuals. We assume that the noise among different players are independent. During the strategy
updating stage at each round of the game, the presence of noise impacts the likelihood of strategy switching for each player according to the
imitation dynamics. To compare with the standard PDG, we still arrange the players on regular lattice networks following Ref. \cite{NM92NATURE}.
We extensively perform the computer simulations to elucidate the impact of different levels of noise. In the remainder of this paper, we first
specify our modified model of PDG; subsequently, we present the main results; and at last, we will summarize the conclusions.

\section{Model}

Let us consider an evolutionary prisoner's dilemma game, where the players occupy the nodes of a regular $L \times L$ square lattice network
with the periodic boundary condition. Each player (each node), e.g. $x$, is initially assigned as either a cooperator ($s_{x}=C$) or a defector
($s_{x}=D$) with the equal probability. In the standard PDG governed by the imitation dynamics, a player updates his strategy according to the
following rules: At each round, the player $x$ performs the PDG with each of his four nearest neighbors, by which he gathers the resulting gains
as his total payoff or fitness. Then he randomly selects one neighboring player $y$, and measures the difference of their fitness at this round,
by which he decides whether changes his own strategy with a probability based the on the Fermi function
\begin{equation}
W(s_y \to s_x)=\frac{1}{1+\exp[(p_x-p_y)/K]},
\end{equation}
where $K$ is the intensity of selection \cite{noi1}, and $p_x, p_y$ are the fitness of players $x,y$, respectively. As mentioned above, in practice,
players may not measure their fitness exactly, due to the presence of noise deriving from many reasons such as the error of observation. Here we
use the parameter $\mu$ to reflect the influence of noise, with $\alpha$ controlling the degree of noise. We define
\begin{equation}
\mu=\alpha\chi,\label{eq.2}
\end{equation}
where $\chi$ is a uniformly distributed random number in the interval [-1,1]. Following the notation suggested in Refs. \cite{spa2,SS09PRE}, we
utilize the rescaled payoff matrix: the temptation to defect $T=b$ (the highest payoff received by a defector playing with a cooperator), reward
for mutual cooperation $R=1$, and both the punishment for mutual defection $P$ and the sucker's payoff $S$ (the lowest payoff required by a
cooperator encountering a defector) is equal to 0, $P=S=0$. The condition $1 \le b \le 2$ ensures the proper payoff ranking. In this modified
PDG model, the dynamics is iterated in accordance with the Monte Carlo (MC) simulation composed of the following elementary steps. First, player
$x$ acquires his total payoff $p_x$ via interacting with his nearest neighbors. With the influence of noise, he evaluates his fitness according
to the relation $F_x=(1+\mu)p_x$. To some extent, this setup is similar to a previously studied model \cite{ag3}. Additionally, compared
with the effects of noise considered in works \cite{noise1,noise2,noise3,noise4}, we mainly consider this factor from the viewpoint of whole
population. Next, by randomly choosing one neighbor $y$ (his fitness $F_y$ is measured in the same way as $x$), the focal
player $x$ adopts $y$'s strategy with the probability
\begin{equation}
W(s_y \to s_x)=\frac{1}{1+\exp[(F_x-F_y)/K]}.
\end{equation}
During one full Monte Carlo step (MCS), each player has a chance to update his strategy according to the above procedure. Starting from a random
initial state, this evolutionary process proceeds until the system reaches the stationary state, at which we record the key statistics: the average
density of cooperator $\rho_C$ in dependence on parameters $K$, $b$, and the connectivity structure.

The simulation results are mainly obtained by implementing the evolutionary dynamics on the square $100\times100$ lattice network. The density of cooperators $\rho_C$ is measured by averaging the last $10^4$ full steps of the overall $2 \times 10^5$ MCS. To overcome the impact of randomness,
final results have been averaged over 40 independent runs for each set of parameters.

\section{Simulation Results}

\begin{figure}[h!]
\begin{center}
\begin{tabular}{c}
\includegraphics[width=10cm]{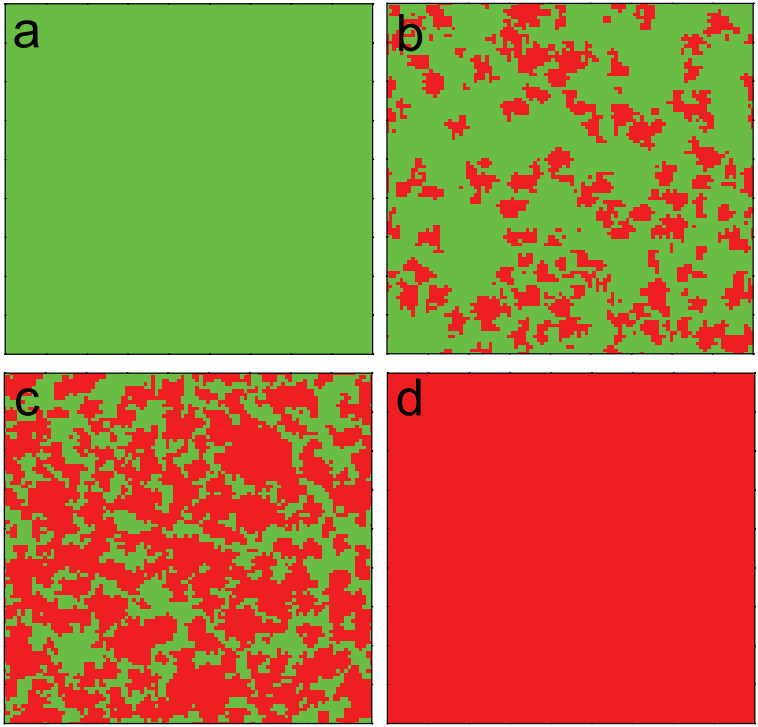}
\end{tabular}
\caption{(\textit{Color Online}) Snapshots of the spatial distribution of cooperators and defectors at the stable states with four
typical values of $\alpha$. Each site corresponds to a player. The cooperators $C$ are represented by the red color, and the defectors
$D$ are denoted by green. From panel (a) to (d), $\alpha=0,0.25,0.5,$ and $0.75$, respectively. We fix $b=1.10$ and $K=0.1$.}\label{fig.1}
\end{center}
\end{figure}

We start by visualizing the spatial distribution of cooperators and defectors at the equilibrium with four typical values
of the parameter $\alpha$. Figure \ref{fig.1} illustrates the results acquired with $b=1.10$, $K=0.1$. As shown in the
upper left panel, the cooperators vanish finally when $\alpha=0$, which conforms to what is expected in the standard model
\cite{NM92NATURE}. When $\alpha>0$, the cooperators begin mushrooming. Strikingly, when $\alpha$ is large enough, the
cooperators will prevail in the system with a negligible number of defectors (see the bottom right panel). This implies
that the cooperation behavior can be remarkably promoted by increasing the value of $\alpha$.

\begin{figure}[h!]
\begin{center}
\begin{tabular}{c}
\includegraphics[width=15cm]{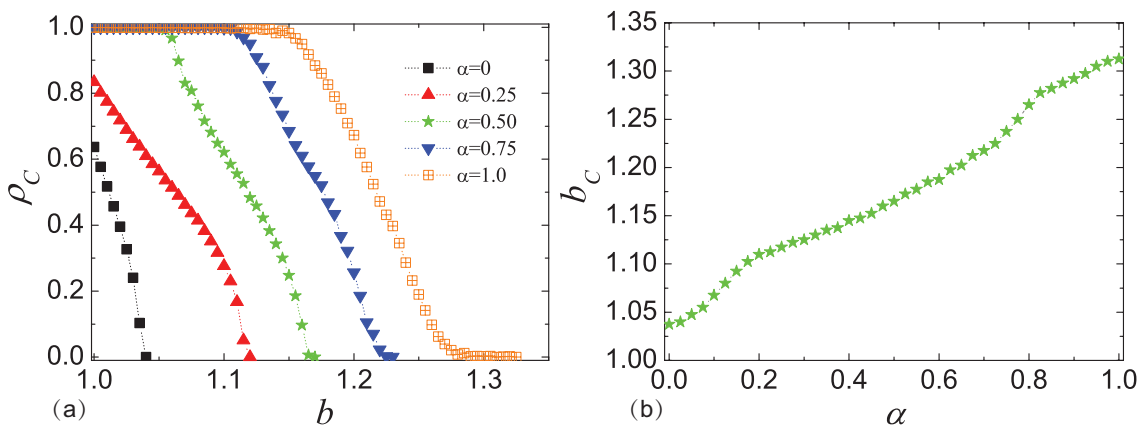}
\end{tabular}
\caption{(\textit{Color Online}) (a) The density of cooperators $\rho_C$ in dependence on the parameters $\alpha$ and $b$ on the regular square
lattice network. (b) Phase diagram of the threshold values of the temptation $b=b_c$, marking the extinction of cooperators if $b>b_c$, as the
parameter $\alpha$ gradually increases. We fix $K=0.1$.}\label{fig.2}
\end{center}
\end{figure}

\begin{figure}[h!]
\begin{center}
\begin{tabular}{c}
\includegraphics[width=9cm]{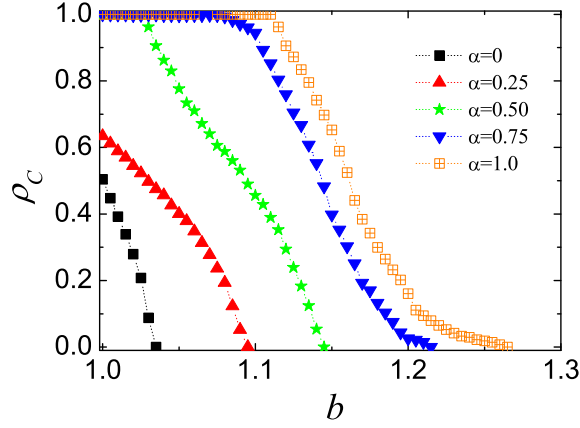}
\end{tabular}
\caption{(\textit{Color Online}) The density of cooperators $\rho_C$ in dependence on the parameter $\alpha$ and $b$ on the triangle
lattice network. We fix $K=0.1$.}
\end{center}\label{fig.3}
\end{figure}

To quantify the impact of each parameter on the behavior of cooperation in detail, we first measure the density of cooperators
$\rho_C$ in dependence on the temptation $b$ for different values of $\alpha$. Figure \ref{fig.2}(a) clearly elucidates the fact
that increasing $\alpha$ promotes the emergence of cooperation. It is worth mentioning that the critical temptation value $b=b_{c}$,
which pinpoints the extinction of cooperators if $b>b_{c}$, increases with the growth of $\alpha$. The monotonous increase of
$b_{c}$ with $\alpha$ is shown in Figure \ref{fig.2}(b). This noise induced enhancement of the survivability of cooperation
on square lattice network raises a question about the universality of this mechanism. Fortunately, the qualitatively consistent
phenomenon is maintained on other types of interaction networks. For instance, in Figure \ref{fig.3}, we report the results on a
triangle lattice graph with several typical values of $\alpha$. It is evident that the positive values of $\alpha$ enhance the
emergence of cooperation. The only difference lies in the fact that the specific values of $b_{c}$ on the triangle lattice graph
are a little smaller than what we can expect on the square lattice network. This implies the potential of the noise-induced
mechanism on promoting the evolution of cooperation.

\vspace{-1cm}

\begin{figure}[h!]
\begin{center}
\begin{tabular}{c}
\includegraphics[width=15cm]{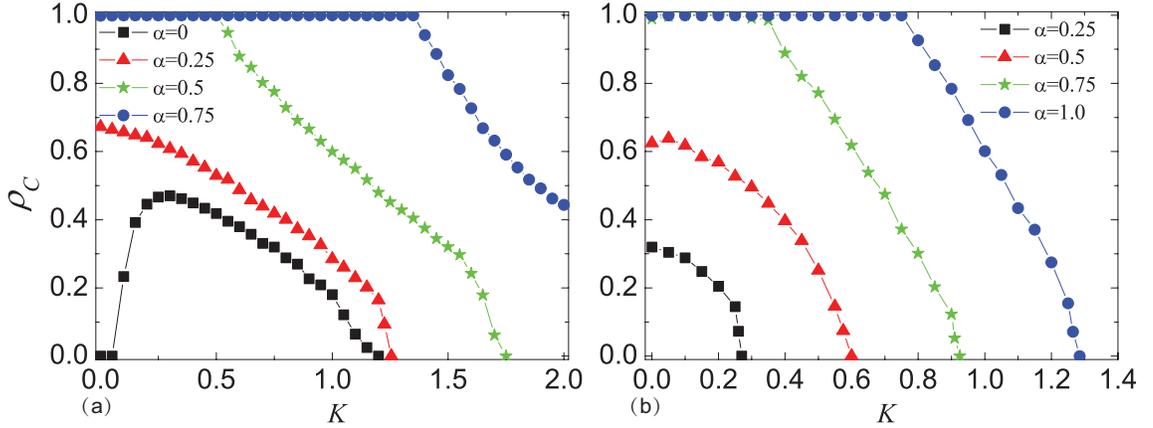}
\end{tabular}
\caption{(\textit{Color Online}) (a) The density of cooperators $\rho_C$ in dependence on the parameters $\alpha$ and $K$ on the
square lattice network for $b=1.06$. (b) The density of cooperators $\rho_C$ in dependence on the parameters $\alpha$ and $K$ on
the square lattice network for $b=1.10$.}\label{fig.4}
\end{center}
\end{figure}

As the form of the Fermi function also allows the players to make irrational decision (e.g., the focal player can adopt the strategy of his
neighbors even if $F_{x}>F_{y}$), we also study the impact of parameter $K$, which characterizes the intensity of selection at the strategy
imitation stage. Figure \ref{fig.4}(a)(b) show the density of cooperators $\rho_C$ in dependence on the parameters $\alpha$ and $K$ on the
regular lattices for $b=1.06$ (a) and $1.10$ (b), respectively. In general, the density of cooperators $\rho_C$ decreases with the augment
of $K$ (except for the scenario with $\alpha=0, b=1.06$ and $K<0.3$, also see \cite{kkk1,kkk2} for comparison). As the increase of $K$ leads to that $\exp[(F_{x}-F_{y})/K]\rightarrow0$,
it is more probable for players to make irrational option, inducing the decrease of $\rho_C$.

\begin{figure}[h!]
\begin{center}
\begin{tabular}{c}
\includegraphics[width=9cm]{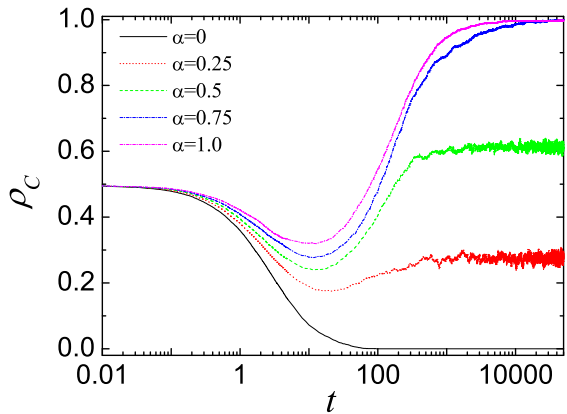}
\end{tabular}
\caption{(\textit{Color Online}) The time courses specifying the evolution of cooperation on the square lattice network with several typical
values of $\alpha$. We fix $b=1.10$ and $K=0.1$.}\label{fig.5}
\end{center}
\end{figure}

To explain the considered mechanism, we report the time courses of $\rho_C$ for several typical values of $\alpha$ on the square lattice
network. Figure \ref{fig.5} illustrates the results obtained with $b=1.10$ and $K=0.1$. The cooperators die out when $\alpha=0$. By gradually
increasing the values of $\alpha$, the stationary state is a mixed $C+D$ phase, where the defectors still occupy a large portion of the network
when $\alpha<0.5$. The cooperators begin prevailing after $\alpha>0.5$, and occupy most nodes of the network after $\alpha\geq0.75$.
Interestingly, as time evolves, there is always an early stage that defectors exploit the ground of cooperators. The smaller the value of
$\alpha$, the more serious the defectors invade at this stage. However, the cooperators will be aroused after this stage, and engage in
beating the opponents.

\begin{figure}[h!]
\begin{center}
\begin{tabular}{c}
\includegraphics[width=10cm]{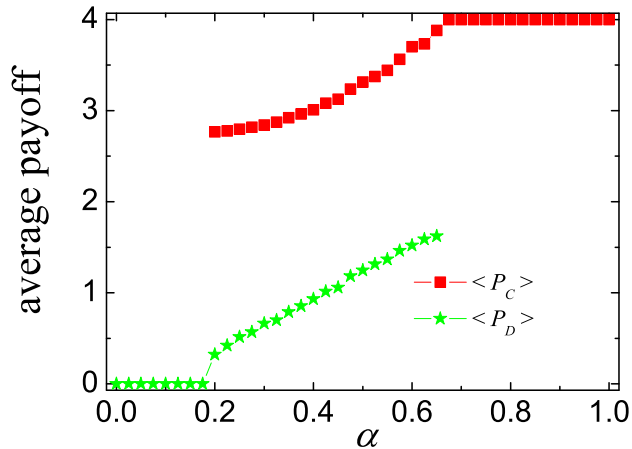}
\end{tabular}
\caption{(\textit{Color Online}) The average payoffs under different values of $\alpha$. $<P_{C}>$ is denoted by red,
$<P_{D}>$ is denoted by green. We fix $b=1.10$ and $K = 0.1$.}
\end{center}\label{fig.6}
\end{figure}

At last, we measure the average payoff of cooperators and defectors on the square lattice network. The total payoff of any player,
e.g., $i$, is the sum after he interacts with all the four neighbors, which is written as
\begin{equation}
p_{i}=\sum_{j\in\Omega_{i}}\phi_{i}^{T}\psi\phi_{j}.
\end{equation}
where $\Omega_{i}$ denotes the set of neighbors of $i$, and $\psi$ is the payoff matrix. $\phi$ refers to the specific strategy
adopted by each player, i.e., $\phi=(1,0)^{T}$ for cooperators and $\phi=(0,1)^{T}$ for defectors. $<P_{C}>$ is defined as the
average payoff of all cooperators, while $<P_{D}>$ is the average payoff of all defectors. As seen in Figure \ref{fig.6}, one can
find that $<P_{C}>$ and $<P_{D}>$ both increase with $\alpha$ when $0.18<\alpha<0.64$. Before $\alpha>0.18$, $<P_{C}>=0$, while
after $\alpha>0.64$, $<P_{D}>=0$. Thus it is clear why the considered mechanism can promote the emergence of cooperation,
although players do encounter the social dilemmas.

\section{Conclusion}

In sum, by introducing the element of noise into the measurement of fitness in the prisoner's dilemma game, our modified model remarkably
enhances the emergence of cooperation. We find that the introduction of noise not only serves as a promising mechanism for the evolutionary
dynamics on the square lattice network but also for that of the triangle lattice graph. Although the defectors can obtain more payoff at the
initial stage, the fast exploitation and thus the shortage of cooperators weakens the advantage of defectors gradually. The remaining
cooperators will form compact clusters preventing the invasion of defectors, and also engage in beating the opponents.

\section*{Acknowledgements}
Gui-Qing Zhang acknowledges partial support from the National Natural Science Foundation of China (Grant No.11247217).
Qi-Bo Sun was supported by the SJTU Student Innovation and Practice Project. Lin Wang also acknowledges the partial
support by Fudan University Excellent Doctoral Research Program (985 Project).




\end{document}